\documentclass[
    10pt,
    twocolumn,
    superscriptaddress,
    nofootinbib,
    amsmath,
    amssymb,
    aps,
    prl
]{revtex4-2}

\usepackage{xcolor}
\usepackage{graphicx}
\usepackage[colorlinks=true, linkcolor=blue, citecolor=red]{hyperref}
\usepackage{amsthm}
\usepackage{array}

\usepackage[usestackEOL]{stackengine}
\usepackage{tikz}
\usetikzlibrary{quantikz}

\newcommand{\mbf}{\mathbf}
\newcommand{\mbb}{\mathbb}
\newcommand{\mc}{\mathcal}
\newcommand{\tr}[1]{\text{Tr}\left[ #1 \right]} 
\newcommand{\ip}[1]{\left\langle #1 \right\rangle} 
\newcommand{\op}[2]{\left|#1\right\rangle\!\left\langle #2\right|} 

\newcommand{\diag}{\text{diag}}

\newcommand{\Net}{\text{Net}}

\newcommand{\rhoNet}{\rho_{[n]}}

\newcommand{\ONet}{O^{\text{\Net}}}

\newcommand{\LNet}{\mc{L}^{\Net}}

\newcommand{\SnBell}{S_{\text{Net}}}

\newcommand{\xv}{\vec{x}}

\newcommand{\Snstar}{S_{n\text{-Star}}}
\newcommand{\Stwostar}{S_{2\text{-Star}}}
\newcommand{\Snstarhat}{\widehat{S}_{n\text{-Star}}}
\newcommand{\Stwostarhat}{\widehat{S}_{2\text{-Star}}}
\newcommand{\SCHSH}{S_{\text{CHSH}}}
\newcommand{\Snchain}{S_{n\text{-Chain}}}
\newcommand{\Snchainhat}{\widehat{S}_{n\text{-Chain}}}

\newcommand{\sigmav}{\vec{\sigma}}

\newcommand{\CNet}{\mbf{C}^{\Net}}
\newcommand{\QNet}{\mc{Q}^{\Net}}

\theoremstyle{definition}
\newtheorem{corollary}{Corollary}

\newtheorem{lemma}{Lemma}

\newtheorem{theorem}{Theorem}

\definecolor{cool_green}{rgb}{0.0, 0.5, 0.0}

\newcolumntype{C}[1]{>{\centering\let\newline\\\arraybackslash\hspace{0pt}}m{#1}}

\begin{document}

\title{Regarding the Maximal Qubit Violations of n-Locality in Star and Chain Networks}

\author{Brian Doolittle}
\affiliation{Department of Physics, University of Illinois at Urbana-Champaign, Urbana, Illinois, 61801, USA}

\author{Eric Chitambar}
\affiliation{Department of Electrical and Computer Engineering, Coordinated Science Laboratory,University of Illinois at Urbana-Champaign, Urbana, Illinois, 61801, USA}

\date{\today}

\begin{abstract}
    The nonlocal correlations of noisy quantum systems are important for both understanding nature and developing quantum technology.
    We consider the correlations of star and chain quantum networks where noisy entanglement sources are measured by nonsignaling parties.
    When pairs of local dichotomic observables are considered on qubit systems, we derive maximal $n$-local violations that are larger and more robust to noise than the maximal $n$-local violations reported previously. To obtain these larger values, we consider observables overlooked in the previous studies.  Thus, we gain new insights into self-testing measurements and entanglement sources in star and chain networks. 
\end{abstract}

\maketitle

\section{Introduction}

Quantum entanglement can be used to generate nonlocal correlations that defy local realism \cite{einstein1935epr, bell1964epr, aspect1981,brunner2014nonlocality}.
In quantum networking \cite{Kimble2008quantum_internet,Simon2017global_quantum_network,wehner2018quantum_internet} , nonlocal correlations have operational advantages in information security \cite{ekert1991_di_qkd,Barrett2005_di_qkd,acin2006_di_crypto,pironio2010di_random_number_generation,liu2018di_rng2,vazirani2019diqkd,Lee2018_di_network,Luo2022_di_network}, multipartite information processing \cite{Brukner2004_quantum_communication_complexity,brassard2005quantum_pseudo_telepathy,Silman2008_nonlocal_games,buhrman2010_communication_complexity,palazuelos2016nonlocal_games_operator_space}, and testing quantum devices \cite{Bardyn2009_di_entangled_state_estimation,Rabelo2011_di_test_entangled_measurements,supic2020_self-test_quantum_system,bancal2018_self-test_network,Renou2018_self-test_network,lee2018_di_nonclassical_measurements,luo2019nonlocal_network_games_self-test,Supic2022_network_self-testing}.
A quantum network consists of many parties linked by entanglement into a graph, leading to novel types of nonlocal correlations \cite{Supic2022_network_self-testing,cavalcanti2011quantum_network_nonlocality,Luo2018_nonlocality_of_networks,renou2019_triangle_network,contreras2021_multipartite_nonlocality,Coiteux-Roy2021_no_bipartite-nonlocal_theory,Pozas2022_full_network_nonlocality,Tavakoli2022_network_nonlocality,lamas2022network_nonlocality}.
We focus on star and chain network topologies because of their important applications in entanglement swapping \cite{Zukowski1993_entanglement_swapping,Bose1998_generalized_entanglement_swapping} and long-distance quantum communication \cite{Bennet1993_teleportation_entanglement,Briegel1998_quantum_repeater,sangouard2011_quantum_repeater}.

We consider the $n$-local scenario where independent entanglement sources link nonsignaling parties.
A network's correlations are $n$-local if they can be obtained using classical sources.
The set of $n$-local correlations is bound by $n$-locality inequalities, whose violation witnesses correlations as non-$n$-local \cite{Tavakoli2022_network_nonlocality}.
Quantum violations of $n$-locality are known for many networks \cite{Branciard_2010_bilocal_correlations,Fritz_2012,Branciard_2012_bilocal_v_nonbilocal,Tavakoli_2014_star,Mukherjee2015chain,chaves_2016_polynomial,Tavakoli2016tree,Rosset_2016_nonlinear_bell_inequalities,Tavakoli_2017_star,Tavakoli2021_EJM,Yang2021_tree_network,Yang2022_tree,yang2022nonlocality_multistar,Mukherjee2022_triangle_network,Bej2022_entanglement_swapping_correlations}, and have been demonstrated experimentally \cite{andreoli2017_bilocal_expt,saunders2017_bilocal_expt,carvacho2017experimental_bilocal,sun2019experimental_bilocal,poderini2020experimental_star_violation,Huang2022_ejm_experimental,Carvacho2022_urban_network_non-n-locality}.

It is crucial to understand how non-$n$-local correlations deteriorate in the presence of noise.
Typically, noise robustness is investigated with respect to a noise model on the sources, communication, or measurement \cite{Branciard_2012_bilocal_v_nonbilocal,Tavakoli_2014_star,Mukherjee2015chain,Gupta2018_amplitude_damping_bilocal,doolittle2022vqo,mukherjee2022persistency}.
However, noise robustness is better characterized by the maximal $n$-local violation attainable for some mixed state.
For instance, the maximal violation of the CHSH inequality \cite{chsh-inequality1969}  is known for any two-qubit mixed state \cite{Horodecki1995}.
Likewise, the maximal $n$-local violations are known in star and chain networks when central parties measure mutually unbiased observables that are separable across qubit systems \cite{gisin2017_bilocal_criterion,andreoli2017maximal_star_violation,kundu2020_nlocal_max_qubit_violations}.
However, variational optimization methods have recently obtained larger $n$-local violations using arbitrary local qubit observables \cite{doolittle2022vqo}.

Our goal is to generalize the $n$-local violations for local qubit observables \cite{doolittle2022vqo}, and to compare them with the maximal violations obtained when the central parties measures their qubits in mutually unbiased bases \cite{gisin2017_bilocal_criterion,andreoli2017maximal_star_violation,kundu2020_nlocal_max_qubit_violations}.
We find that reference \cite{Tavakoli_2017_star} misses an equality condition between the star network's $n$-local violation and its upper bound, the geometric mean of each source's CHSH violation.
Exploiting this condition, we achieve the upper bound for all two-qubit mixed states where the external parties must measure observables in mutually unbiased bases, \textit{e.g.}, $\sigma_x$ and $\sigma_z$.
These $n$-local violations are larger and more robust to noise than those obtained previously under the assumption that the central parties, instead, measure observables in mutually unbiased bases.

In this paper, we first formalize star and chain quantum networks and their non-$n$-local correlations.
We then present our results regarding the maximal qubit violations of $n$-locality in star and chain networks.
Finally, we discuss their application in self-testing, network nonlocality, and maximizing $n$-local violations.

\section{Methods}

\subsection{Star and Chain Quantum Networks}

\begin{figure}[t]
    \centering
    \includegraphics[width=.48\textwidth]{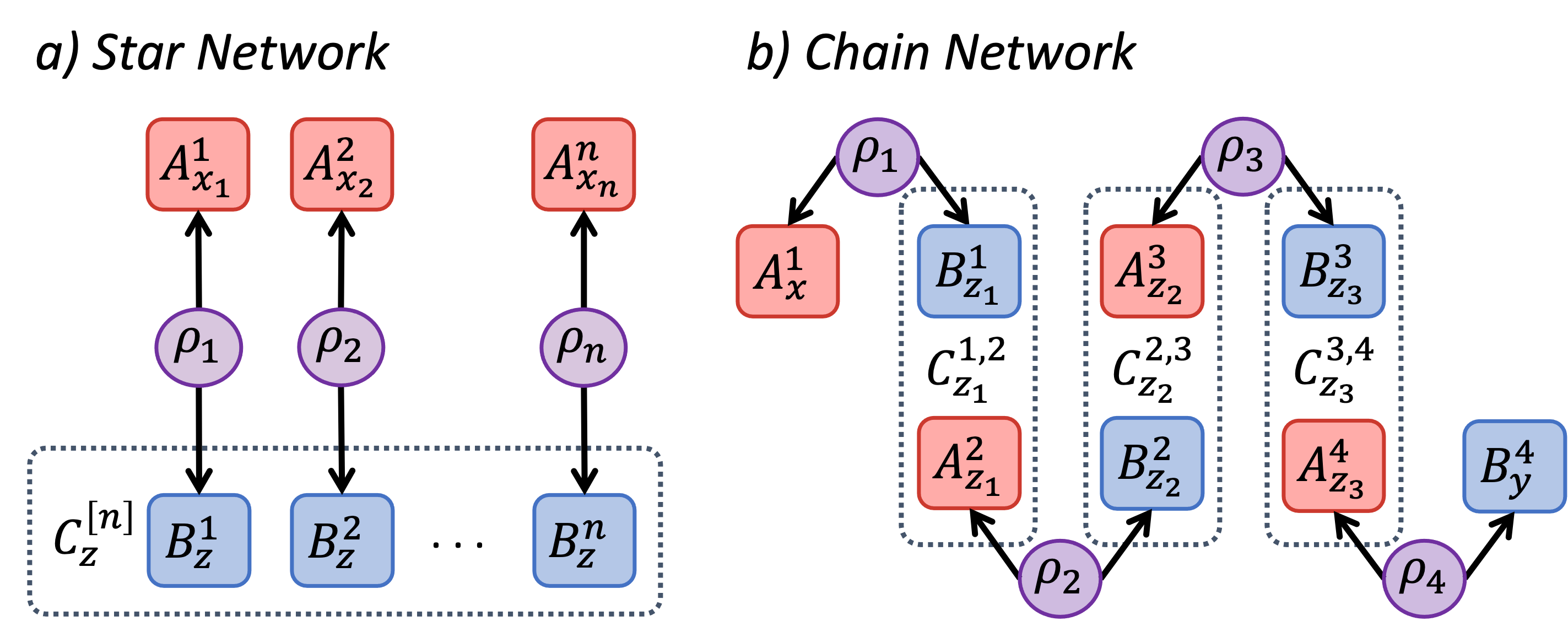}
    \caption{
        Sources (ellipses) and parties (rectangles).
        Parties $A_{x_i}^i$ and $B^i_{y_i}$ measure qubit observables.
        The dashed rectangles group multi-qubit observables.
    }
    \label{fig:n-local_networks_diagram}
\end{figure}

Star and chain quantum networks each have $n$ entanglement sources that link $m=n+1$ dichotomic parties into their respective topology (see Fig.~\ref{fig:n-local_networks_diagram}).
In aggregate, the network prepares the state $\rhoNet \equiv \bigotimes_{i=1}^n \rho_i$ where each source emits a two-qubit mixed state $\rho_i \in D(\mc{H}^{A_i}\otimes \mc{H}^{B_i})$.
Each party measures their local qubits using a Hermitian observable that has $\pm 1$ eigenvalues and the dichotomic observable is conditioned upon a private binary input, \textit{e.g.}, $x_j,y_j$, or $z_j\in\mbb{B}\equiv\{0,1\}$.

Similarly to previous works \cite{gisin2017_bilocal_criterion,andreoli2017maximal_star_violation,kundu2020_nlocal_max_qubit_violations}, we consider local qubit observables.
That is, the state $\rho_i\in D(\mc{H}^{A_i}\otimes \mc{H}^{B_i})$ is measured by the qubit observables, $A^i_{x_i} = \vec{\alpha}^{i}_{x_i}\cdot\sigmav$ and $B^i_{y_i}=\vec{\beta}^{i}_{y_i} \cdot \sigmav$, where $|\vec{\alpha}^i_{x_i}|=|\vec{\beta}^i_{y_i}|=1$, $\sigmav = (\sigma_x,\sigma_y,\sigma_z$),  and $\langle A^i_{x_i}\otimes B^i_{y_i}\rangle_{\rho_i} = \tr{A^i_{x_i}\otimes B^i_{y_i} \rho_i}$ is the expected parity of the two-qubit measurement result.
A multi-qubit observable is then the product of  qubit observables, \textit{e.g.}, $C^{L_j}_{z_j} = \bigotimes_{i\in L_j} B^i_{z_j}$, where $L_j\subseteq [n]$ indexes the linked sources and the $\pm 1$ eigenvalues of $C^{L_j}_{z_j}$ correspond to the parity of the $|L_j|$-qubit measurement result.

Since multi-qubit observables are separable, the network's observable is as well, $O^{\text{Net}}_{\xv} \equiv (\bigotimes_{i=1}^n A^{i}_{x_i}\otimes B^i_{y_i})$, where $\xv \in\mbb{B}^m$ is the $m$-bit string containing each party's input and, $x_i,y_i\in\mbb{B}$ are set to their corresponding party's input as shown in Fig.~\ref{fig:n-local_networks_diagram}.
The observable $O^{\text{Net}}_{\xv}$ is dichotomic and measures the parity of the network's $m$-bit output.
Hence, the network correlator factors across across all two-qubit pairs
\begin{align}
    \ip{O^{\text{Net}}_{\xv}}_{\rhoNet} 
    & = \prod_{i=1}^n \ip{A^i_{x_i}\otimes B^i_{y_i}}_{\rho_i}, \label{eq:n-local_correlator_factorization}
\end{align}
where the correlator gives the expected parity of all qubit measurements in the network.
We then characterize the network by its correlations, $\CNet \equiv \{ \langle O^{\Net}_{\xv} \rangle_{\rhoNet} \}_{\xv\in\mbb{B}^m}$.

\subsection{Quantum Non-n-Locality}

We compare the correlations of quantum and classical networks.
We use $\QNet$ to denote the set of quantum network correlations where $\CNet\in\QNet$ if the network correlator $\langle \ONet_{\xv} \rangle_{\rhoNet}$ factors as in Eq.~\eqref{eq:n-local_correlator_factorization}.
We use $\LNet$ to denote the set of $n$-local network correlations where $\CNet \in \LNet$ if all sources emit classical states, \textit{i.e.}, each $\rho_i\in D(\mc{H}^{A_i}\otimes\mc{H}^{B_i})$ is diagonal in the computational basis \cite{Tavakoli2022_network_nonlocality}. 
In general $\LNet\subseteq \QNet$, and the quantum network correlations $\CNet\in \QNet$  are non-$n$-local if $\CNet\not\in\LNet$.
Non-$n$-local correlations violate some $n$-locality inequality that  bounds $\LNet$  \cite{brunner2014nonlocality,Tavakoli2022_network_nonlocality}.
We denote these inequalities as $\SnBell(\mbf{C}^{\Net}) \leq \beta$ where $\beta$ is the $n$-local bound and $\SnBell(\mbf{C}^{\Net})$ is some nonlinear function referred to as the $n$-local network score.
For a given state $\rhoNet$, the maximal $n$-local network score is
\begin{equation}
    \SnBell^{\star}(\rhoNet) \equiv  \max_{\{\ONet_{\xv}\}_{\xv\in\mbb{B}}} \SnBell(\CNet)
\end{equation}
where $\CNet = \{\tr{\ONet_{\xv}\rhoNet}\}_{\xv\in\mbb{B}^m}$ and we optimize over all local qubit observables $\ONet_{\xv} = \bigotimes_{i=1}^n (A^i_{x_i}\otimes B^i_{y_i})$.

The nonlocal content of a two-qubit mixed state $\rho_i$ is found in its correlation matrix, $T_{\rho_i}\in\mbb{R}^{3\times 3}$, that has elements $T_{\rho_i}^{(k,\ell)} = \tr{\sigma_k \otimes \sigma_\ell \rho_i}$ $\forall$ $k,\ell\in\{x,y,z\}$.
For an arbitrary mixed state $\tilde{\rho}_i\in D(\mc{H}^{A_i}\otimes \mc{H}^{B_i})$, the correlation matrix can be diagonalized as \cite{Gamel2016_twoqubit_state_space}
\begin{equation}\label{eq:correlation_matrix_diagonal}
    T_{\rho_i} = \diag(\vec{\tau}_i) = R^{A_i} T_{\tilde{\rho}_i} (R^{B_i})^T
\end{equation}
where $R^{A_i},R^{B_i}\in SO(3)$ and $\vec{\tau}_i \equiv (\tau_{i,1},\tau_{i,2},\tau_{i,0})$ where $1\geq\tau_{i,0}\geq\tau_{i,1}\geq|\tau_{i,2}|\geq0$ are the singular values of $T_{\tilde{\rho}_i}$.
If $R^{A_i}$ and $R^{B_i}$ are chosen freely, the values $\tau_{i,j}\in\vec{\tau}_i$ can be permuted and/or sign-flipped as long as the constraint holds that $\det(T_{\tilde{\rho}_i})=\det(T_{\rho_i}) = \prod_{j=1}^3 \tau_{i,j}$.
Since a homomorphism maps $SU(2)$ to $SO(3)$ \cite{cornwall1984group}, the rotations, $R^{A_i}$ and $R^{B_i}$, correspond to qubit unitaries, $V^{A_i}$ and $V^{B_i}\in SU(2)$, such that Eq.~\eqref{eq:correlation_matrix_diagonal} becomes $\rho_i = V^{A_i}\otimes V^{B_i} \tilde{\rho}_i (V^{A_i}\otimes V^{B_i})^\dagger$.
Therefore, we assume without loss of generality, that $\rho_i$ satisfies $T_{\rho_i} = \diag(\vec{\tau}_i)$ because we give measurements local qubit unitary freedom.

As an example, consider the case where one source links two parties, $A_x$ and $B_y$.
All local correlations $\mbf{C}^\Net\in\mc{L}^{\Net}$ satisfy the CHSH inequality \cite{chsh-inequality1969}
\begin{align}
    \SCHSH(\mbf{C}^{\Net}) \equiv& \textstyle\sum_{y \in \mbb{B}}\ip{O^{AB}_y} \leq 2 \label{eq:chsh_inequality}\\
    \text{where}\quad  O^{AB}_y = (&A_0 + (-1)^y A_1)\otimes B_y. \label{eq:CHSH_pm_observables}
\end{align}
The CHSH score quantifies the performance in a game where the goal is to maximize the likelihood that the binary inputs and outputs satisfy $a\oplus b = x\wedge y$ \cite{brunner2014nonlocality}.

All quantum correlations $\CNet\in\QNet$ are bounded as $\SCHSH(\CNet)\leq 2\sqrt{2}$ \cite{Cirelson1980} and, for any two-qubit mixed state $\rho$, the maximal CHSH score is \cite{Horodecki1995}
\begin{equation}\label{eq:max_CHSH_score}
    \SCHSH^\star(\rho) = 2\sqrt{\tau^2_0 + \tau^2_1}
\end{equation}
where $\tau_j$ are the singular values of $T_\rho$.
Note that $\rho$ can be used to generate nonlocal correlations if and only if $\tau^2_0 + \tau^2_1 > 1$.
Hence we define a classical state $\gamma$ where $T_\gamma = \diag(0,0,1)$ such that $\SCHSH^\star(\gamma) = 2$.
Examples include the product of two pure qubit states, $\gamma=\op{\psi}{\psi}\otimes\op{\phi}{\phi}$, or a shared coin flip, $\gamma = \frac{1}{2}(\op{00}{00} + \op{11}{11})$.

To obtain $\SCHSH^\star(\rho)$, there are two choices of optimal qubit observables,
\begin{align}
    \{&A_x = (1-x)\sigma_z + (x) \sigma_x,\;B_y = \textstyle\frac{\tau_0 \sigma_z + \tau_1 (-1)^{y} \sigma_x}{\sqrt{\tau^2_0 + \tau^2_1}}\}\; \text{or} \label{eq:optimal_CHSH_observables}\\
    &\{\hat{A}_x = \textstyle\frac{\tau_0 \sigma_z + \tau_1 (-1)^{x} \sigma_x}{\sqrt{\tau^2_0 + \tau^2_1}},\;\widehat{B}_y = (1-y)\sigma_z + (y) \sigma_x\},\label{eq:optimal_CHSH_observables2}
\end{align}
where each choice yields a distinct expectation
\begin{align}
    \ip{O^{AB}_y}_{\rho} = \textstyle \sqrt{\tau_0^2 + \tau_1^2} = \textstyle \frac{1}{2}\SCHSH^\star(\rho) \label{eq:optimal_chsh_pm_observable2}, \\
    \langle O^{\hat{A}\widehat{B}}_y\rangle_{\rho} = \textstyle\frac{2}{\sqrt{\tau^2_0 + \tau^2_1}} \big( \tau^2_0 (1 -y) +  \tau^2_1 y\big).\label{eq:optimal_chsh_pm_observable}
\end{align}
To achieve $\SCHSH^\star(\rho)$, one party must measure observables in mutually unbiased bases, \textit{e.g.}, $\sigma_x$ and $\sigma_z$, while the other party's observables might be nonorthogonal, \textit{e.g.}, $|\tr{B_0 B_1}| \geq 0$.
However, Eqs.~\eqref{eq:optimal_CHSH_observables} and~\eqref{eq:optimal_CHSH_observables2} show that either party can perform either of these measurements.
A fact that follows from $T_\rho$ being symmetric and the two parties being indistinguishable.
From a self-testing perspective, this symmetry implies that a maximal CHSH score, $\SCHSH^\star(\rho)$, is insufficient on its own to determine with certainty which of the two parties measured in mutually unbiased bases.
As we will see, this symmetry is broken in the star and chain networks.

\section{Results}

\subsection{Maximal Qubit Violations of Star n-Locality}

Consider the star network as depicted in Fig.~\ref{fig:n-local_networks_diagram}.a).
All $n$-local correlations $\CNet\in\LNet$ satisfy the $n$-locality inequality \cite{Tavakoli_2014_star}
\begin{align}
    \Snstar(\mbf{C}^{\Net}) &\equiv \sum_{z\in\mbb{B}}\left|I_{n,z}\left(\mbf{C}^{\Net}\right)\right|^{\frac{1}{n}} \leq 1, \;\;\text{where}\label{eq:n-local_star_inequality} \\
    I_{n,z}\left(\mbf{C}^{\Net}\right) \equiv& \textstyle\frac{1}{2^n} \sum_{\xv \in \mbb{B}^m} (-1)^{z\bigoplus_{j=1}^m x_j}\langle O^{\text{Star}}_{\xv}\rangle_{\rhoNet}, \label{eq:star_correlator_combo}\end{align}
$O^{\text{Star}}_{\xv} = (\bigotimes_{i=1}^n A^i_{x_i})\otimes C^{[n]}_{z}$, and $C^{[n]}_z = \bigotimes_{i=1}^n B^i_z$.
Note that $S_{1\text{-Star}}(\mbf{C}^{\Net})= \frac{1}{2}\SCHSH(\mbf{C}^{\Net})$ \cite{chsh-inequality1969}, and that $\Stwostar(\CNet)$ is the bilocal score \cite{Branciard_2010_bilocal_correlations, Branciard_2012_bilocal_v_nonbilocal}.
Considering local qubit observables, the network correlator factors as in Eq.~\eqref{eq:n-local_correlator_factorization} and Eq.~\eqref{eq:n-local_star_inequality} becomes \cite{Tavakoli_2017_star}
\begin{align}
    \Snstar(\CNet) &= \textstyle \frac{1}{2}{\textstyle\sum}_{z\in\mbb{B}}\big|{\textstyle\prod}_{i=1}^n \ip{O^{A_i B_i}_z}_{\rho_i}\big|^{\frac{1}{n}} \label{eq:factored_star_inequalilty}
\end{align}
where $O^{A_i B_i}_{z}$ is the CHSH observable in Eq.~\eqref{eq:CHSH_pm_observables}.
Since each party measures the parity of their local qubits, Eq.~\eqref{eq:factored_star_inequalilty} quantifies the likelihood that $\bigoplus_{i=1}^n (a_i\oplus b_i) = \bigoplus_{i=1}^n( x_i\wedge z)$, which is the $XOR$ of CHSH games played using each independent source.

It follows that for any quantum correlations $\CNet\in\QNet$, the $n$-local star score is bounded by the geometric mean of independent CHSH scores \cite{Tavakoli_2014_star,Tavakoli_2017_star},
\begin{equation}\label{eq:upper_bound_to_max_star_violation}
    \Snstar(\CNet) \leq \textstyle\frac{1}{2} \textstyle\prod_{i=1}^n \SCHSH^\star(\rho_i)^{\frac{1}{n}},
\end{equation}
which holds for entangled states of any dimension \cite{Munshi2022_star_bound_dimension,kumar2022_star_bound_dimension}.

\begin{lemma}\label{lemma:star_optimality_conditions}
    Equality holds in Eq.~\eqref{eq:upper_bound_to_max_star_violation} if:
    \begin{enumerate}
        \item $\langle O^{A_i B_i}_{z=0}\rangle_{\rho_i} = \langle O^{A_i B_i}_{z=1}\rangle_{\rho_i} \; \forall \; i\in [n]$
        \item $\langle O^{A_1 B_1}_{z}\rangle_{\rho_1} =\dots = \langle O^{A_n B_n}_{z}\rangle_{\rho_n}\; \forall\; z\in\mbb{B}$ \cite{Tavakoli_2017_star}
        \item $\langle O^{A_i B_i}_{z=0}\rangle_{\rho_i} = \langle O^{A_i B_i}_{z=1}\rangle_{\rho_i} = 0$ for some $i\in [n]$
    \end{enumerate}
    
    \begin{proof}
        Let $X_z^i = \frac{1}{2}\langle O^{A_i B_i}_z\rangle_{\rho_i}$, then Eq.~\eqref{eq:upper_bound_to_max_star_violation} becomes $\sum_{z\in\mbb{B}}\left(\prod_{i=1}^n X_z^i\right)^{1/n} \leq  \prod_{i=1}^n (\sum_{z\in\mbb{B}} X_z^i)^{1/n}$.
        Equality holds when: 1.~$X_0^i=X_1^i$ $\forall$ $i\in[n]$, 2.~$X_z^1=\dots=X_z^n$ $\forall$ $z\in\mbb{B}$, or 3.~$X_0^1=X_1^1 = 0$.
    \end{proof}
\end{lemma}

Reference \cite{Tavakoli_2017_star} presents a similar result to Lemma~\ref{lemma:star_optimality_conditions}, however, Condition 2 is an if and only if statement, while Conditions~1 and~3 are overlooked.

\begin{theorem}\label{thm:max-star-violation}
    For any ensemble of two-qubit mixed states $\rhoNet$, the maximal $n$-local star score in Eq.~\eqref{eq:upper_bound_to_max_star_violation} can be obtained using local qubit observables.
    That is,
    \begin{equation}\label{eq:maximal_n-local_star_score}
        \Snstar^\star(\rhoNet) = \textstyle\frac{1}{2}\textstyle\prod_{i=1}^n\SCHSH^\star(\rho_i)^{\frac{1}{n}}.
    \end{equation}
    \begin{proof}
        For all $i\in[n]$, consider $A^i_{x_i}$ and $B^i_{y_i}$ to be the optimal CHSH observables from Eq.~\eqref{eq:optimal_CHSH_observables} that obtain the expectation in Eq.~\eqref{eq:optimal_chsh_pm_observable2}.
        Since Condition~1 of Lemma~\ref{lemma:star_optimality_conditions} is satisfied, the maximal $n$-local star score is obtained.
    \end{proof}
\end{theorem}

Theorem~\ref{thm:max-star-violation} contrasts with previous results, which assume that the multi-qubit observables are in the mutually unbiased bases, 
$\widehat{C}^{[n]}_z = (1-z)\bigotimes_{i=1}^n \sigma^{B_i}_z + z \bigotimes_{i=1}^n \sigma^{B_i}_x$.
Then, the maximal $n$-local star score becomes \cite{gisin2017_bilocal_criterion,andreoli2017maximal_star_violation,kundu2020_nlocal_max_qubit_violations}
\begin{align}
    \Snstarhat^\star(\rhoNet) = \sqrt{{\textstyle \prod}_{i=1}^n \tau^{2/n}_{i,0} + {\textstyle \prod}_{i=1}^n \tau^{2/n}_{i,1}} \label{eq:max_star_violation_literature}
\end{align}
where the external parties measure the qubit observables
\begin{align}
    \hat{A}^i_{x_i} &= \frac{\prod_{i=1}^n \tau^{1/n}_{i,0}\sigma_z + (-1)^{x_i}\prod_{i=1}^n \tau^{1/n}_{i,1}\sigma_x}{\sqrt{{\prod_{i=1}^n \tau_{i,0}^{2/n} + \prod_{i=1}^n \tau_{i,1}^{2/n}}}}.\label{eq:optimal_star_observable_literature}
\end{align}

\begin{corollary}\label{cor:star_equality}
    $\Snstar^\star(\rhoNet) = \Snstarhat^\star(\rhoNet)$ if: 1.) $\tau_{i,0}=\tau_{i,1}$ $\forall$ $i\in[n]$, 2.) $\tau_{1,z}=\dots=\tau_{n,z}$ $\forall$ $z\in\mbb{B}$, or 3.) $\tau_{i,0}=\tau_{i,1}=0$ for some $i\in[n]$.
\end{corollary}
In general $\Snstarhat^\star(\rhoNet) \leq \Snstar^\star(\rhoNet)$ where equality occurs in the special cases described in Corollary~\ref{cor:star_equality}.
In practice, these conditions are often assumed to hold.
For example, Condition 1 holds for white noise is models and Condition 2 holds if all sources emit the same state.

We note that $\Snstarhat^\star(\rhoNet)$ is maximal under the assumption that the \textit{central} party measures the observables $\widehat{C}^{[n]}_z\in\{\bigotimes_{i=1}^n \sigma_x,\;\bigotimes_{i=1}^n \sigma_z\}$, which are in mutually unbiased bases.
Otherwise, if parties measure arbitrary local qubit observables, then $\Snstar^\star(\rhoNet)$ is maximal and the \textit{external} parties each must measure their qubit observables in mutually unbiased bases, \textit{e.g.}, $A^i_{x_i}\in\{\sigma_x,\sigma_z\}$.
As noted earlier, a violation $\SCHSH^\star(\rho) > 2$ is alone insufficient to determine which qubit was measured in mutually unbiased bases.
In the star network, if $\Snstar^\star(\rhoNet) > \Snstarhat^\star(\rhoNet)$ holds for a known state $\rhoNet$, then the external parties measured observables $A^i_{x_i}\in\{\sigma_x,\sigma_z\}$.
This fact  could be used to self-test that qubits are being measured in mutually unbiased bases. 

In most cases, $\Snstar^\star(\rhoNet) > \Snstarhat^\star(\rhoNet)$. 
Reference \cite{doolittle2022vqo} notes an extreme case, in which $1\leq k < n$ sources each emit the  classical state $\gamma_i=\op{00}{00}$ while the remaining sources each emit maximally entangled states.
In this case, $\Snstar^\star(\rhoNet) = 2^{(n-k)/2n}>1 = \Snstarhat^\star(\rhoNet)$, thus showing an example where $n$-local violations are found when previous results predict that no violation can occur.
We now use Theorem~\ref{thm:max-star-violation} to generalize this example.
\begin{corollary}\label{cor:star_separation_partially_classical}
    Let $\gamma_{[k]} = \bigotimes_{i=1}^k \gamma_i$ where $T_{\gamma_i}=\diag(0,0,1)$, and $\rho_{[k+1,n]} = \bigotimes_{i=k+1}^n \rho_i$ for any $\rho_i\in D(\mc{H}^{A_i}\otimes \mc{H}^{B_i})$.
    Then, $\Snstarhat^\star(\gamma_{[k]}\otimes\rho_{[k+1,n]}) \leq 1$, and $\Snstar^\star(\gamma_{[k]}\otimes\rho_{[k+1,n]}) = S_{(n-k)\text{-Star}}^\star(\rho_{[k+1,n]})^{(n-k)/n}$.
\end{corollary}

Using Corollary~\ref{cor:star_separation_partially_classical} we find that, if $k=1$, the maximal $n$-local violation is bounded as $\Snstar^\star(\rhoNet) \leq 2^{(n-1)/2n}$.
Thus, if $\Snstar(\CNet) >  2^{(n-1)/2n}$, then all sources are nonclassical.
Similarly, if the central party measures the observable $\widehat{C}^{[n]}_{z}$ such that $\Snstarhat^\star(\rhoNet)$ is maximal, then the violation $\Snstarhat^\star(\rhoNet) > 1$, witnesses all sources to be nonclassical.
Therefore, in either case, a sufficiently large $n$-local violation asserts that no classical sources are present.
Indeed, Theorem~\ref{thm:max-star-violation} leads to bounds whose violation witnesses \textit{full quantum network nonlocality} where all sources are required to be nonclassical \cite{Pozas2022_full_network_nonlocality}.

\begin{figure}[t]
    \centering
    \includegraphics[width=.48\textwidth]{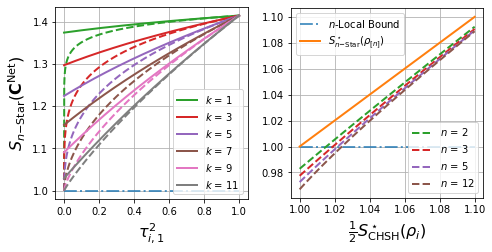}
    \caption{
        We compare $\Snstar^\star(\rhoNet)$ (solid) with $\Snstarhat^\star(\rhoNet)$ (dashed).
        (Left) We consider $k<n=12$ noisy sources that have $\tau_{i,0}=1$ and $\tau_{i,1}\in[0,1]$. The remaining $(n-k)$ sources have $\tau_{i,0}=\tau_{i,1}=1$.
        (Right) For all $i\in[n]$, we set $\frac{1}{2}\SCHSH^\star(\rho_i)=\beta^\star\in[1.0,1.1]$  while $\tau^2_{i,1} = (\beta^\star)^2 - \tau^2_{i,0}$ where $\tau_{i,0}^2$ are evenly spaced in $\big[\frac{1}{2}(\tau^2_{i,0}+\tau^2_{i,1}),\min\{1,\tau^2_{i,0}+\tau^2_{i,1} \}\big]$.
    }
    \label{fig:star_network_numerics}
\end{figure}

Overall, the $n$-local violations of $\Snstar^\star(\rhoNet)$ are more robustness to noise than $\Snstarhat^\star(\rhoNet)$. 
In Fig.~\ref{fig:star_network_numerics}, we illustrate the separation between $\Snstar^\star(\rhoNet)$ and $\Snstarhat^\star(\rhoNet)$.
In the left plot, we consider $k$ sources to be affected by colored noise that dampens $\tau_{i,1}$ but preserves $\tau_{i,0}$, while the remaining sources are noiseless.
As $\tau_{i,1}$ becomes small, a large separation exists.
In the right plot, we consider a case where Corollary~\ref{cor:star_equality} does not hold.
That is, for all $i\in[n]$, $\SCHSH^\star(\rho_i)$ is constant, but the pair $\tau_{i,0}$ and $\tau_{i,1}$ are unique.
We thus build examples where $\Snstar^\star(\rhoNet)>1>\Snstarhat^\star(\rhoNet)$. 

\subsection{Maximal Qubit Violations of Chain n-Locality}

Consider the $n$-local chain network as depicted in Fig.~\ref{fig:n-local_networks_diagram}.b).
All network correlations $\CNet\in\LNet$ satisfy the $n$-local chain inequality \cite{Mukherjee2015chain}
\begin{align}
    \Snchain(\mbf{C}^{\Net}) &= \sum_{y\in\mbb{B}}\left|J_{n,z}\left(\mbf{C}^{\Net}\right)\right|^{\frac{1}{2}} \leq 1 \quad \text{where} \label{eq:n-local_chain_inequality}\\
    J_{n,z}\left(\mbf{C}^{\Net}\right) &= {\textstyle\frac{1}{4}}\sum_{x,y \in \mbb{B}} (-1)^{z(x + y)}\ip{O^{\text{Chain}}_{x, y,z}},\label{eq:n-local_chain_correlators}
\end{align}
$O^{\text{Chain}}_{x, y, z} = A^1_{x}\otimes(\bigotimes_{i=1}^{n-1} C^{\:i,i+1}_z ) \otimes B^n_{y}$, and $C^{\:i,i+1}_z = B^i_{z}\otimes A^{i+1}_z$.
Considering local qubit observables, we use Eq.~\eqref{eq:n-local_correlator_factorization} to rewrite Eq.~\eqref{eq:n-local_chain_inequality} and Eq.~\eqref{eq:n-local_chain_correlators} as
\begin{align}
    &\Snchain(\CNet)=\Stwostar(\CNet_{1,n})\textstyle\prod_{i=2}^{n-1} \ip{A^i_z\otimes B^i_z}_{\rho_i}\label{eq:qubit_chain_factored} \\
    &J_{n,z}(\CNet)=\prod_{i\in\{1,n\}}\frac{1}{2} \langle O^{A_i B_1}_z \rangle_{\rho_i}\prod_{i=2}^{n-1} \ip{A^i_z\otimes B^i_z}_{\rho_i}\label{eq:optimal_chain_J_ny}
\end{align}
where $O^{A_iB_i}_z$ is the CHSH observable in Eq.~\eqref{eq:CHSH_pm_observables} and $\prod_{i\in\{1,n\}}\frac{1}{2}\langle O^{A_i B_i}_z \rangle_{\rho_i} = I_{2,z}(\CNet_{1,n})$ as in Eq.~\eqref{eq:star_correlator_combo} where $\CNet_{1,n}$ denotes the correlations of sources $1$ and $n$.

All quantum correlations $\CNet\in\QNet$ are bounded as $\Snchain(\mbf{C}^{\Net}) \leq \sqrt{2}$ \cite{Mukherjee2015chain}.
We improve this bound as
\begin{equation}\label{eq:tight_chain_bound}
    \Snchain(\CNet)\leq \Stwostar^\star(\rho_1\otimes\rho_n)\textstyle\prod_{i=2}^{n-1}\sqrt{\tau_{i,0}}
\end{equation}
where we used the fact that $|\langle A^i_z \otimes B^i_z\rangle| \leq \tau_{i,0}$.

\begin{theorem}\label{thm:max_chain_violation}
    For any ensemble of two-qubit mixed states $\rhoNet$, the maximal $n$-local chain score in Eq.~\eqref{eq:tight_chain_bound} is achieved using local qubit observables.
    That is,
    \begin{equation}\label{eq:maximal_chain_violation}
        \Snchain^\star(\rhoNet) = \Stwostar^\star(\rho_1\otimes\rho_n){\textstyle \prod}_{i=2}^{n-1} \sqrt{\tau_{i,0}}.
    \end{equation}
    \begin{proof}
        Since we consider local qubit observables, we define the optimal observables with respect to the source they measure.
        For all $i\in [2,n-1]$, let $\rho_i$ be measured by the observable $A^i_0\otimes B^i_0= A^i_1 \otimes B^i_1= \sigma_z\otimes \sigma_z$ such that $\langle A^i_z \otimes B^i_z\rangle_{\rho_i} = \tau_{i,0}$.
        Then for $i\in\{1,n\}$, let the external parties measure $A^1_0 = B^n_0 = \sigma_z$ and $A^1_1 = B^n_1 = \sigma_x$, while $B^1_{z} = (\tau_{1,0} \sigma_z + \tau_{1,1}(-1)^{z} \sigma_x)/(\tau_{1,0}^2 + \tau_{1,1}^2)^{1/2}$ and $A^n_{z} = (\tau_{n,0} \sigma_z + \tau_{n,1}(-1)^z \sigma_x)/(\tau_{n,0}^2 + \tau_{n,1}^2)^{1/2}$.
        Inserting these observables into Eq.~\eqref{eq:qubit_chain_factored} yields the upper bound on the $n$-local chain score in Eq.~\eqref{eq:tight_chain_bound}.
    \end{proof}
\end{theorem}

The maximal $n$-local chain score $\Snchain^\star(\rhoNet)$ derived in Theorem~\ref{thm:max_chain_violation} contrasts with the maximal $n$-local violations derived in reference \cite{kundu2020_nlocal_max_qubit_violations}.
In this work, the central parties are assumed to measure the observables $\widehat{C}^{i,i+1}_z = (1-z)\sigma_z\otimes \sigma_z +z\sigma_x\otimes \sigma_x$, leading to the maximal $n$-local chain score \cite{kundu2020_nlocal_max_qubit_violations}
\begin{align}
    \Snchainhat^\star(\rhoNet) &\equiv \sqrt{\textstyle\prod_{i=1}^n \tau_{i,0} + \textstyle\prod_{i=1}^n \tau_{i,1}}
    \label{eq:max_chain_violation_literature}
\end{align}
where the two external parties measure the observables
\begin{align}
    \hat{A}^1_{x} =  \frac{\prod_{i=1}^n \sqrt{\tau_{i,0}}\sigma_z + (-1)^{x}\prod_{i=1}^n \sqrt{\tau_{i,1}}\sigma_x}{\sqrt{{\prod_{i=1}^n \tau_{i,0} + \prod_{i=1}^n \tau_{i,1}}}}, \label{eq:optimal_chain_observable_literature}
\end{align}
and similarly for $\widehat{B}^n_y$.
In general, $\Snchain^\star(\rhoNet)\geq \Snchainhat^\star(\rhoNet)$ where equality occurs only in special cases.
\begin{corollary}\label{cor:chain_equality}
    $\Snchain^\star(\rhoNet) = \Snchainhat^\star(\rhoNet)$ if $\tau_{i,0}=\tau_{i,1}$ for all $i\in[2,n-1]$ and $\Stwostar^\star(\rho_1\otimes \rho_n)=\Stwostarhat^\star(\rho_1\otimes\rho_n)$.
\end{corollary}
Theorem~\ref{thm:max_chain_violation} shows that it is sufficient measure all sources $i\in[2,n-1]$ using the observable $\sigma_z\otimes\sigma_z$.
Thus, the maximal $n$-local chain score can be achieved when all sources $i\in[2,n-1]$ are classical.
This fact was observed in reference \cite{doolittle2022vqo}.
For instance, let $\rhoNet = \rho_1\otimes (\bigotimes_{i=2}^{n-1} \gamma_i)\otimes \rho_n$ where $T_{\gamma_i}=\diag(0,0,1)$, then $\Snchain^\star(\rhoNet) = \Stwostar^\star(\rho_1\otimes\rho_n)$ and $\Snchainhat^\star(\rhoNet)\leq 1$.
However, if the central parties measure the observables $\widehat{C}^{(i,i+1)}_z$, then an $n$-local violation $\Snchainhat(\CNet)>1$ asserts that each source is nonclassical.
As a consequence, full quantum network nonlocality \cite{Pozas2022_full_network_nonlocality} cannot be witnessed in the chain network with respect to local qubit observables, but can if central parties measure their qubits in mutually unbiased bases.

\begin{figure}[t]
    \centering
    \includegraphics[width=.48\textwidth]{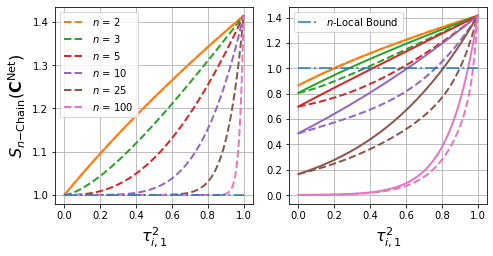}
    \caption{
        For various $n$, $\Snchain^\star(\rhoNet)$ (solid) is compared with $\Snchainhat^\star(\rhoNet)$ (dashed).
        (Left) We set $\tau^2_{i,0}=1$ and vary $\tau^2_{i,1}\in[0,1]$.
        (Right) We set $\tau^2_{i,0}= \frac{3}{4} + \frac{1}{4}\tau^2_{i,1}$ and vary $\tau^2_{i,1}\in[0,1]$.
    }
    \label{fig:chain_network_numerics}
\end{figure}

Overall, the $n$-local violations of $\Snchain^\star(\rhoNet)$ are more robust to noise than the $n$-local violations of $\Snchainhat^\star(\rhoNet)$. 
In Fig.~\ref{fig:chain_network_numerics}, we illustrate cases where $\Snchain^\star(\rhoNet)>\Snchainhat^\star(\rhoNet)$.
In both plots, we consider uniform noise on all sources such that Corollary~\ref{cor:star_equality} is satisfied such that $\widehat{S}_{2\text{-Star}}^\star(\rho_1\otimes\rho_n) = \Stwostar^\star(\rho_1\otimes \rho_n)$.
Thus, the separation between $\Snchain^\star(\rhoNet)$ and $\Snchainhat^\star(\rhoNet)$ is inherent to the chain network.
In the left plot, we consider colored noise where $\tau_{i,0}$ is preserved but $\tau_{i,1}$ is damped.
For all $n > 2$, we find $\Snchain^\star(\rhoNet) = \Stwostar^\star(\rho_1\otimes\rho_n)\geq\Snchainhat^\star(\rhoNet)\geq 1$ where the separation increases with $n$.
In the right plot, we add white noise such that $\tau_{i,0}$ is also damped, but the bias $\tau_{i,0} > \tau_{i,1}$ is preserved.
For all $n>2$, we find that $\Stwostar^\star(\rho_1\otimes\rho_n)>\Snchain^\star(\rhoNet) \geq \Snchainhat^\star(\rhoNet)$ and we find examples where $\Snchain^\star(\rhoNet) > 1 > \Snchainhat^\star(\rhoNet)$.

\section{Discussion}

In this work, we investigate the non-$n$-local correlations that can form in noisy star and chain quantum networks.
In Theorems~\ref{thm:max-star-violation} and~\ref{thm:max_chain_violation}, we derive the maximal $n$-local star and chain scores, $\Snstar^\star$ and $\Snchain^\star$, for any ensemble of two-qubit mixed states measured using local qubit observables.
These $n$-local network scores are larger and more robust to noise than $\Snstarhat^\star$ and $\Snchainhat^\star$, which are maximal under the assumption that multi-qubit measurements are in mutually unbiased bases that are separable across qubit systems  \cite{andreoli2017maximal_star_violation,kundu2020_nlocal_max_qubit_violations}.

The optimal observables for these distinct $n$-local violations might be useful for testing the sources and measurements of star and chain networks.
In particular, we note an asymmetry where a maximal $n$-local violation satisfying $\Snstar^\star(\rhoNet)> \Snstarhat^\star(\rhoNet)$ requires the external parties to apply qubit observables in mutually unbiased bases, $\sigma_x$ and $\sigma_z$.
A fact that could be used in self-testing measurements.
Furthermore, our results relate to the framework of full network nonlocality \cite{Pozas2022_full_network_nonlocality} where we obtain bounds whose violation indicates that no classical sources are present.
Thus, a sufficiently large $n$-local violation asserts that all sources are nonclassical.

An advantage arises when optimizing a network's observables for the maximal $n$-local violation given uncharacterized sources.
Namely, to achieve $\Snstarhat^\star$ and $\Snchainhat^\star$, the qubit observables of external parties depend upon the states emitted from all sources, as shown in Eq.~\eqref{eq:optimal_star_observable_literature} and Eq.~\eqref{eq:optimal_chain_observable_literature}.
Thus, when optimizing a single qubit observable, the network must be considered as a whole.
In contrast, to achieve $\Snstar^\star$ and $\Snchain^\star$, the optimal qubit observables depend only on the state they measure, allowing the observables on each source to be optimized as independent CHSH violations.
This distinction simplifies the practical task of optimizing nonlocal correlations on quantum hardware \cite{Suprano2021,Poderini2022_black-box,doolittle2022vqo}.

Although $\Snstarhat^\star$ and $\Snchainhat^\star$ are only maximal in special cases, we find that most results obtained using these quantities hold.
Mainly, either the central parties measure in mutually unbiased bases, or Corollaries~\ref{cor:star_equality} or~\ref{cor:chain_equality} hold due to the presence of uniform state preparations or white noise.
However, the fact that $\Snstar^\star(\rhoNet)\geq \Snstarhat^\star(\rhoNet)$ and $\Snchain^\star(\rhoNet)\geq \Snchainhat^\star(\rhoNet)$ is important to networking applications in information security \cite{Lee2018_di_network,Luo2022_di_network}, self-testing \cite{bancal2018_self-test_network,Renou2018_self-test_network,lee2018_di_nonclassical_measurements,luo2019nonlocal_network_games_self-test,Supic2022_network_self-testing}, nonlocality sharing \cite{mahato2022_nonlocality_sharing,zhang2022nonlocality_sharing,mao2022recycling_nonlocality_sharing,wang2022network_sharing}, and quantum steering \cite{jiang2022quantum_steering_star_network}.
In future works, it would be interesting to see if the maximal $n$-local violations can be improved when parties have more inputs or outputs, or for topologies beyond stars and chains.
Furthermore, we expect that the variational quantum optimization techniques for quantum networks \cite{doolittle2022vqo} can be applied more broadly to obtain further insights.

\subsection*{Acknowledgements}

This work was supported by NSF Award DMR-1747426 and NSF Award 2016136.

\bibliography{references}

\end{document}